\documentclass[prb,aps,twocolumn,superscriptaddress,showpacs]{revtex4-1}

\usepackage{graphicx}
\usepackage{epstopdf}

\begin{document}

\title{Competing spin density wave, collinear, and helical magnetism in Fe$_{1+x}$Te}

\author{C. Stock}
\affiliation{School of Physics and Astronomy, University of Edinburgh, Edinburgh EH9 3JZ, UK}
\author{E. E. Rodriguez}
\affiliation{Department of Chemistry of Biochemistry, University of Maryland, College Park, MD, 20742, U.S.A.}
\author{P. Bourges}
\affiliation{Laboratoire Leon Brillouin, CEA-CNRS, Universite Paris-Saclay, CEA Saclay, 91191 Gif-sur-Yvette Cedex, France}
\author{R. A. Ewings}
\affiliation{ISIS Facility, Rutherford Appleton Laboratory, Didcot, OX11 0QX, UK}
\author{H. Cao}
\affiliation{Quantum Condensed Matter Division, Oak Ridge National Laboratory, Oak Ridge, Tennesee 37831, USA}
\author{S. Chi}
\affiliation{Quantum Condensed Matter Division, Oak Ridge National Laboratory, Oak Ridge, Tennesee 37831, USA}
\author{J. A. Rodriguez-Rivera}
\affiliation{NIST Center for Neutron Research, National Institute of Standards and Technology, 100 Bureau Drive, Gaithersburg, Maryland, 20899, USA}
\affiliation{Department of Materials Science, University of Maryland, College Park, Maryland 20742, USA}
\author{M. A. Green}
\affiliation{School of Physical Sciences, University of Kent, Canterbury, CT2 7NH, UK}

\date{\today}

\begin{abstract} 

The Fe$_{1+x}$Te phase diagram consists of two distinct magnetic structures with collinear order present at low interstitial iron concentrations and a helical phase at large values of $x$ with these phases separated by a Lifshitz point.  We use unpolarized single crystal diffraction to confirm the helical phase for large interstitial iron concentrations and polarized single crystal diffraction to demonstrate the collinear order for the iron deficient side of the Fe$_{1+x}$Te phase diagram.  Polarized neutron inelastic scattering show that the fluctuations associated with this collinear order are predominately transverse at low energy transfers, consistent with a localized magnetic moment picture.  We then apply neutron inelastic scattering and polarization analysis to investigate the dynamics and structure near the boundary between collinear and helical order in the Fe$_{1+x}$Te phase diagram.  We first show that the phase separating collinear and helical order is characterized by a spin-density wave with a single propagation wave vector of ($\sim$ 0.45, 0, 0.5).  We do not observe harmonics or the presence of a charge density wave.  The magnetic fluctuations associated with this wavevector are different from the collinear phase being strongly longitudinal in nature and correlated anisotropically in the (H,K) plane.   The excitations preserve the $C_{4}$ symmetry of the lattice, but display different widths in momentum along the two tetragonal directions at low energy transfers.  While the low energy excitations and minimal magnetic phase diagram can be understood in terms of localized interactions, we suggest that the presence of density wave phase implies the importance of electronic and orbital properties.
 
\end{abstract}


\maketitle

\section{Introduction}

The discovery of unconventional superconductivity in the iron based pnictides~\cite{Kamihara08:130} and chalcogenides~\cite{Hsu08:105}, and the subsequent materials effort, has lead to the founding of a number of magnetic iron-based materials which are proximate to superconductivity, or strongly correlated electronic phases.~\cite{Dai15:87,Inosov16:17,Ishida09:78,Green10:6,Johnston10:59,Stock16:28,Wen11:74,Dai12:8,Lumsden10:22}  However, identifying and understanding the parent phases of these systems remains an unresolved challenge.  Unlike the case of the cuprate high temperature superconductors~\cite{Birgeneau06:75,Kastner98:70}, where a Mott insulating phase is parent to high temperature superconductivity, parent phases of iron based supercondutors appear to be semi or poorly metallic and it remains unclear if these parent compounds are based on localized magnetic order or more metallic, itinerant behavior.  Also, unlike in the cuprates which appear to derive from a single electronic band, multi band~\cite{Raghu08:77,Graser09:11,Lei12:85} and orbital effects~\cite{Kruger09:79,Lv010:82} appear to be required to understand the electronics and magnetism of iron based systems.  

The single layered chalcogenide superconducting system Fe$_{1+x}$Te$_{1-y}Q_{y}$ (where $Q$=Se or S) is important owing to its relatively simple crystal structure based on a single FeTe layer~\cite{Fruchart75:10,Hsu08:105,Sales09:79,Deguchi12:13,Rossler16:84,Mizuguchi10:79} and also because it displays a strongly localized electronic character in comparison to other iron based systems.~\cite{Yin11:10}    Fe$_{1+x}$Te$_{1-y}Q_{y}$ also allows two chemical variables to control magnetic, structural, and electronic properties: $x$ represents the amount of interstitial iron disordered throughout the crystal, and $y$ the amount of anion substitution.    However, several studies have found that these two variables ($x$ and $y$) are correlated and both are central in determining superconducting properties.~\cite{McQueen09:79,Stock12:85,Vivanco16:242}  In particular, structurally, the role of the tetrahedral bond angles have been identified as being tuned with interstitial iron.~\cite{Rodriguez11:2,Xu16:93} Magnetically, interstitial iron has been implicated as the origin of several doping induced magnetic and structural phase transitions.~\cite{Tang16:6,Thampy12:108}   Because of this strong correlation between anion and interstitial iron doping, it is important to understand the parent Fe$_{1+x}$Te phase diagram where a single variable is tuned.

\begin{figure}[t]
\includegraphics[width=8.8cm] {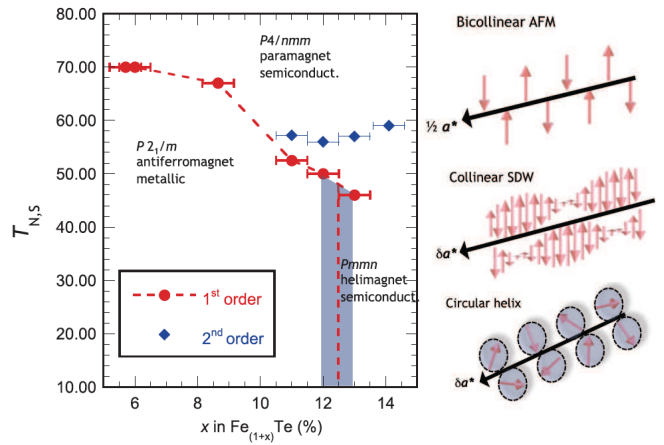}
\caption{\label{structures} The magnetic, structural, and electronic phase diagram of Fe$_{1+x}$Te taken from Refs. \onlinecite{Rodriguez13:88,Rodriguez11:84} based on neutron and x-ray diffraction and resistivity data.  A schematic of the different magnetic structures are also shown.}
\end{figure}

The combined magnetic and structural phase diagram for Fe$_{1+x}$Te is illustrated in Fig. \ref{structures} (taken from Refs. \onlinecite{Rodriguez13:88,Rodriguez11:84,Koz13:88}).   The phase diagram is divided into two key sections by the concentration $x\sim$ 0.12.  For low interstitial iron concentrations, a commensurate and collinear magnetic phase is present characterized by a ``double-stripe" structure with magnetic moments aligned along the $b$ axis and magnetic Bragg peaks at (0.5, 0, 0.5) or ($\pi$, 0).~\cite{Bao09:102}  The transition to this collinear magnetic phase as a function of temperature is first order~\cite{Li09:79} and accompanied by a transition from a semi/poorly metallic state at high temperatures to a metallic phase at low temperature.~\cite{Chen09:15,Rodriguez13:88}  The second region is at large interstitial iron concentrations where helical magnetic order is present and also characterized by a second order phase transition.  The resistivity in this region of the phase diagram is semi/poorly metallic at all temperatures.    Structurally, the low-$x$ collinear magnetic phase is characterized by a low temperature monoclinic unit cell ($P2_{1}/m$) where the helical high-$x$ phase has an orthorhombic unit cell ($Pmmm$) at low temperatures.  The order parameters and critical scattering was investigated in Ref. \onlinecite{Rodriguez13:88} and the it was concluded (based on critical exponents) that the order parameters were decoupled in these two extremes of the interstitial iron phase diagram.

The point separating this line of first and second order transitions is defined as a Lifshitz point.~\cite{Collins:book}  The magnetic phase near this point has been investigated in several studies~\cite{Rodriguez11:84,Rodriguez13:88} and is unusual in several regards in comparison to the two extreme phases discussed above.  First, while the wavevector characterizing magnetic order near this point is incommensurate at ($\sim$ 0.45, 0, 0.5), the magnetic structure is collinear as proven through polarized neutron scattering with the magnetic moments aligned along the $b$ axis.  Second, the phase is long-ranged (determined by the resolution of the neutron diffractometer) along $c$, but short-range along $a$.  Third, the critical exponents defining the structural and  magnetic phase transitions were the same within error suggesting the structural and magnetic order parameters were coupled.  This phase was recently the subject of a theoretical study~\cite{Materne15:115} where it was suggested the origin was based upon localized topological defects (termed ``Solitonic spin-liquid") derived from a combined Mossbauer and theoretical analysis.   The localized topological defects have parallels with proposed structures for the spin-glass phase in the lamellar cuprates.~\cite{Hasselmann04:69}

Perhaps most relevant to superconductivity, the magnetic phase associated with this incommensurate wave vector has been observed to compete and even coexist with superconducting phases in anion doped materials.   Temperature dependent studies of the magnetic fluctuations in Fe$_{1+x}$Se$_{y}$Te$_{1-y}$ have observed static magnetic correlations peaked at wavevectors less than the commensurate H=0.5 position.~\cite{Xu10:82}  Though interpreted as commensurate correlations, the data reported in Figure 3 of Ref. \onlinecite{Xu10:82} is consistent with short range incommensurate correlations.  Incommensurate correlations near the superconducting transition as a function of Se doping were also identified in Ref. \onlinecite{Bendele10:82}.  More correlated, in momentum, incommensurate scattering at (0.46, 0, 0.5) was reported in superconducting samples of Fe$_{1.02}$Te$_{0.75}$Se$_{0.25}$.~\cite{Wen09:80}   Studies of Fe$_{1-z}$Cu$_{z}$Te  observe that this short range incommensurate order at ($\sim$ 0.42, 0, 0.5) seems to be stabilized with copper doping.~\cite{Wen12:86}  All of these studies illustrate that short-range incommensurate order at ($\sim$ 0.45, 0, 0.5) competes and even coexists with superconductivity and also the two distinct collinear and helical magnetic phases described above in the absence of anion doping.

A key underlying question surrounding this is whether the ground state of the parent phase Fe$_{1+x}$Te can be understood in terms of an itinerant metal~\cite{Mazin10:464,Hirayama13:87,Akbari11:84,Han09:103,Singh10:104,Han10:104,Ding13:87,Chubukov16:93} or whether a localized magnetic state based on Heisenberg interactions through the Te $5p$ band~\cite{Ma009:102,Ducatman12:109,Chen13:88,Hu12:85} is more relevant.  Itinerant models would point to density wave phases with the length of the magnetic spin varying while localized models would propose spatially localized defects and competitions between localized structures where the length of the spin is preserved.   Initial studies on superconducting Fe$_{1+x}$Te$_{0.6}$Se$_{0.4}$ suggested the importance of itinerant effects~\cite{Argyriou10:81,Lumsden10:6} however neutron inelastic scattering have more recently been interpreted in terms of a localized model~\cite{Lipscombe11:106} with competing spin states.~\cite{Zal11:107}   Other high energy neutron inelastic scattering work has observed an hour glass type of dispersion with spectral weight extending up to $\sim$ 200 meV and the total integrated spectral weight be short of expectations based on a purely localized picture.~\cite{Stock14:90}

Here, we investigate the question of itinerant vs electronic effects near the Lifshitz point in Fe$_{1+x}$Te using a combination of neutron elastic and inelastic scattering.  We investigate the magnetic fluctuations in the collinear phase and show that the low-energy fluctuations are primarily transverse indicating a strong localized character at low-energies.  For large interstitial iron concentrations, we investigate the magnetic structure using single crystals and confirm the helical phase and also the magnetic structure of the interstitial sites.  We then investigate the incommensurate short-range order in single crystals for $x \sim$ 0.12, at the boundary between collinear and helical magnetic phases. 

\section{Experimental}

The single crystals discussed here are the same samples used in previous studies where the preparation techniques are outlined in detail.~\cite{Rodriguez13:88,Rodriguez11:84}  The interstitial iron concentration was determined with single crystal x-ray and also powder neutron diffraction as discussed previously.~\cite{Rodriguez11:84}  

The magnetic fluctuations in the collinear magnetic phase were studied on single crystals of Fe$_{1.057(7)}$Te.  The temperature dependence of the magnetic fluctuations was investigated using the MAPS chopper spectrometer (ISIS, UK).  The sample was aligned such that Bragg positions of the form (H0L) lay within the horizontal scattering plane with the $c$-axis aligned parallel to the incident beam.  The $t_{0}$ chopper was spun at a frequency of 50 Hz and phased to remove high energy neutrons from the target.  A ``sloppy" Fermi Chopper was used to monochromate the incident beam with E$_{i}$=75 meV, and the Fermi chopper was spun at a frequency of 200 Hz (with an elastic energy resolution of 4.0 meV at full width half maximum).  The sample was cooled with a bottom loading closed cycle refrigerator. 

As noted previously and discussed in the supplementary information of Ref. \onlinecite{Stock14:90}, with the $c$ axis parallel to the incident beam $\vec{k}_{i}$, the H and K axes are projected onto the MAPS detectors providing a good experimental configuration to measure the momentum dependence in this plane.  However, the value of L, or the projection along $c$, changes as a function of energy transfer and also coordinates (H,K).  This has been discussed in previous works on cuprate superconductors and used in the case of the bilayer YBa$_{2}$Cu$_{3}$O$_{6.5}$ to extract magnetic optic and acoustic fluctuations.~\cite{Stock05:71}   With our configuration of E$_{i}$=75 meV, (H,K,L)=(0.5,0.5,$\sim$0.5) is found at 8-10 meV, near the peak in the magnetic intensity measured with a triple axis spectrometer where all three parameters are determined.  

For studying the helical magnetic structure in single crystals of Fe$_{1.141(5)}$Te, we utilized the HB-3A four-circle diffractometer at the High-flux Isotope Reactor (HFIR) at Oak Ridge National Laboratory (Oak Ridge, USA).  Both nuclear and magnetic reflections were measured using the Si(220) monochromator with a wavelength of 1.5424 \AA.  

To study the short-range incommensurate collinear magnetic phase we investigated single crystals of Fe$_{1.124(5)}$Te.  Neutron diffraction measurements were performed on the Wide Angle Neutron Diffractometer (WAND) located at the HIFR Reactor (Oak Ridge, USA).   WAND is configured in an energy integrating ``2-axis" mode with E$_{i}$=36.3 meV using Ge(113).  An oscillating collimator is used after the detector to limit backgrouund.   Further studies searching for harmonics and also establishing the magnetic wavevector were performed at the MACS cold triple-axis spectrometer at NIST (Gaithersburg, USA).~\cite{Rodriguez08:19}  The final energy was fixed at E$_{f}$=3.6 meV and the elastic scattering plane was measured using the 20 double-bounce PG(002) analyzing crystals and detectors.  Each detector channel was collimated using 90$'$ Soller slits before the analyzing crystal.  The sample was aligned in the (H0L) scattering plane and cooled in a closed cycle refrigerator (WAND) and orange-cryostat (MACS).

Polarized neutron inelastic scattering~\cite{Shirane:book} was performed on the 4F1 triple-axis spectrometer located at the LLB (Saclay, France) to study the magnetic fluctuations in Fe$_{1.057(7)}$Te.  The incident beam was polarized with a supermirror and analyzed with a Heusler crystal,  A Beryllium filter was used on the scattered side to remove higher order contamination of the beam.  The final energy was fixed to E$_{f}$=5.0 meV (with an energy resolution of 0.23 meV, full width at half maximum).  The measured flipping ratio, with the sample, of 8 is significantly reduced owing to the presence of ferromagnetic iron on the surface of the sample which depolarizes the neutron beam.  For this reason, polarized neutron experiments were unsuccessful for high interstitial iron concentrations, where a helical magnetic structure is found, owing to the large amount of ferromagnetic iron on the surface of the sample.

\section{x=0.057(7) - Collinear magnetism and stripy fluctuations}

We first discuss the temperature dependent magnetic dynamics in the collinear phase of the Fe$_{1+x}$Te phase diagram by studying single crystals of Fe$_{1.057(7)}$Te.  Fe$_{1.057(7)}$Te is placed on the iron deficient side of the phase diagram shown in Fig. \ref{structures} and has a first order transition at 75 K to a collinear magnetic phase accompanied by a structural transition from a tetragonal (space group $P4/nmm$) to monoclinic (space group $P2_{1}/m$) unit cell.  We first show how the magnetic fluctuations change in the (H,K) plane as a function of temperature and then study the anisotropy of these temperature dependent fluctuations using polarized neutrons.  

\begin{figure}[t]
\includegraphics[width=8.8cm] {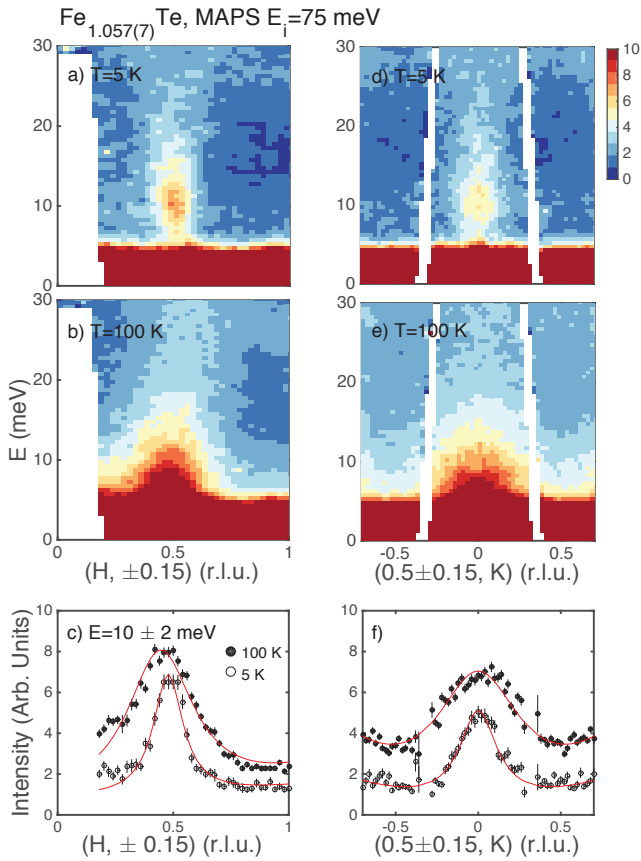}
\caption{\label{constQ} Constant momentum slices taken on the MAPS spectrometer with E$_{i}$=75 meV on Fe$_{1.057(7)}$Te.  $(a-c)$ show scans and one dimensional cuts along the H direction and $(d-f)$ show scans along the K direction at 5 and 100 K.  The constant energy cuts in panels $(c)$ and $(f)$ were done at 10 $\pm$ 2 meV.}
\end{figure}

\begin{figure}[t]
\includegraphics[width=9.5cm] {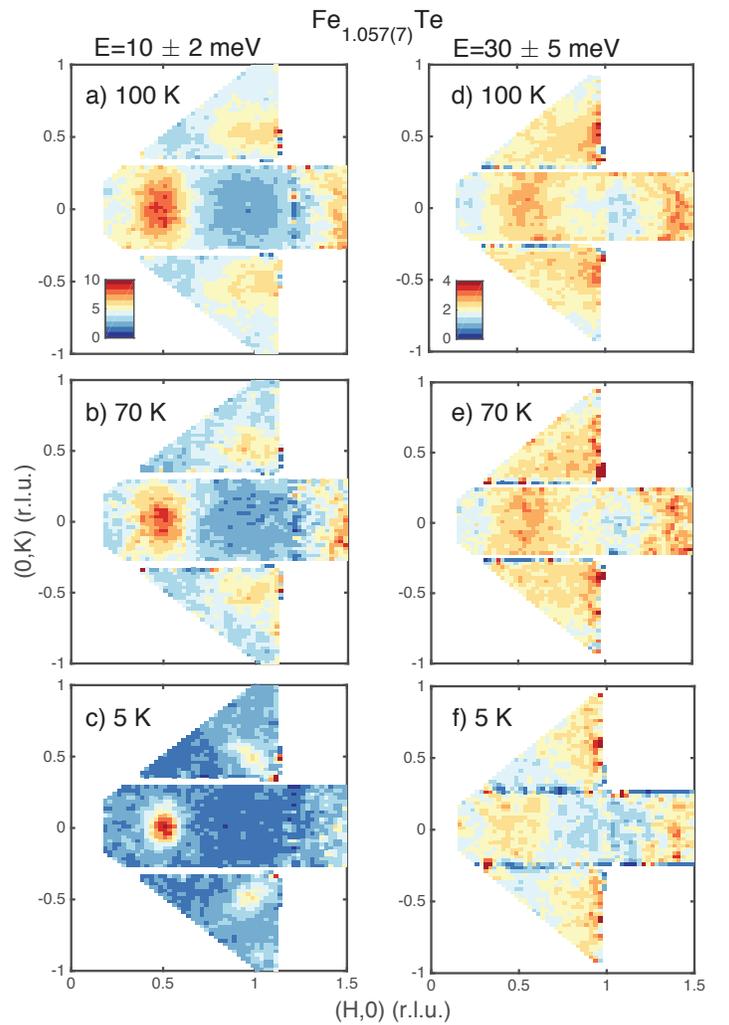}
\caption{\label{constE} Constant energy slices taken on the MAPS spectrometer with E$_{i}$=75 meV and k$_{i}$ aligned along the $c$ axis.  $(a-c)$ displays slices at E=10 $\pm$ 2 meV and $(d-f)$ illustrate scans at 30 $\pm$ 5 meV.  Temperatures of 5, 70, and 100 K are shown for each energy transfer.}
\end{figure}

Figure \ref{constQ} shows constant momentum slices taken near (H,K)=(0.5,0) at 5 K, below the transition to collinear magnetic order, both along the H and K directions taken on the MAPS chopper spectrometer with E$_{i}$=75 meV.  Unlike the case of Se doped Fe$_{1+x}$Te$_{1-y}$Se$_{y}$ where the magnetic fluctuations are peaked near the ($\pi$, $\pi$) position, in parent Fe$_{1+x}$Te, the magnetic correlations are peaked near ($\pi$, 0).~\cite{Chi11:84}  As previously published, the low temperature magnetic fluctuations are strongly correlated along both the $a$ and $b$ directions and become one dimensional at higher energy transfers in excess of $\sim$ 30 meV.~\cite{Stock14:90}    This is confirmed in the constant momentum slices in panels $(a)$ and $(d)$ and the corresponding cuts in panels $(c)$ and $(f)$ where the magnetic fluctuations are strongly correlated in momentum along both the H and K directions at E=10 meV.  A considerable broadening occurs at high temperatures of 100 K, however, the fluctuations remain anisotropic in momentum at this temperature as illustrated in constant momentum slices in panels $(b)$ and $(e)$ and also cuts $(c)$ and $(f)$ taken at E=10 $\pm$ 2 meV.

Figure \ref{constE} displays constant energy slices at low energy transfers of 10 $\pm$ 2 meV (panesl $a-c$) and also 30 $\pm$ 5 meV (panels $d-f$).  The data are also from the MAPS spectrometer with E$_{i}$=75 meV.  At low temperatures and low energies displayed in panel $(c)$, a constant energy map shows that the scattering is well correlated in both the $a$ and $b$ directions.  At 70 K (panel $b$) close to the first order transition to collinear order, the results discussed above is further confirmed showing broadened, yet still anisotropic correlations.  At 100 K (well above T$_{N}$), however as illustrated in panel $(a)$, the scattering becomes more isotropic being broader along $b$ yet there is still a clear anisotropy in the correlations along $a$ and $b$.  At higher energies (E=30 $\pm$ 5 meV displayed in panels $d-f$), a different picture emerges with the magnetic fluctuations being more elongated along the K direction at 5 K indicative of one dimensional fluctuations.  At higher temperatures of 70 K, the magnetic correlations become isotropic along the H and K directions with the scattering forming nearly a ring in momentum at 100 K.

As noted previously in a high energy neutron scattering study~\cite{Stock14:90} as a function of interstitial iron concentration, the magnetic excitations extend up to at $\sim$ 200 meV and this is also confirmed by two-magnon results using Raman.~\cite{Okazaki11:83}  Within error of $\pm$ 15 \%, we observe no temperature dependence to the integrated intensity at 5, 70, and 100 K integrating over energy transfers up to 50 meV.  While the analysis is sensitive to how the elastic line is treated, the increase in spectral weight in the inelastic channel is accounted for by the loss of spectral weight at the magnetic Bragg position within error.   This contrasts with some previous studies on Fe$_{1+x}$Te (Ref.\onlinecite{Zal11:107}), however we emphasize that our measurements are performed on a different sample which is located at a different point in the magnetic and structural phase diagram drawn in Fig. \ref{structures}.  We have also discussed possible sources of error due to low-energy phonons in the supplementary information in Ref. \onlinecite{Stock14:90}.  In the collinear phase of Fe$_{1.057(7)}$Te, we therefore do not observe evidence of a spin transition, but rather a re-distribution of spectral weight from the elastic line to the inelastic position and also throughout the Brillouin zone as a function of temperature. 

The constant energy and momentum cuts in Figs. \ref{constQ} and \ref{constE} illustrate that the fluctuations become considerably broadened in momentum and energy crossing the Neel transition (T$_{N}$=75 K).  Fig. \ref{constQ} panels $(c)$ and $(f)$ show that the magnetic fluctuations remain peaked around K=0 and H=0.5, however at high temperatures of 100 K above the first order magnetic and structural transition, the magnetic fluctuations at 10 meV are slightly displaced in H to lower values away from the commensurate H=0.5 position.   The nature of these incommensurate fluctuations will be discussed in more detail below.  It is interesting to note that while the magnetic fluctuations become considerably broadened at high temperatures, they do remain very anisotropic in the (H,K) plane as illustrated in Fig. \ref{constE} panel $(a)$ which is at 100 K, well above the Neel transition temperature.  Gaussian fits to the data produce an anisotropy in momentum with widths of $\xi_{a}$/$\xi_{b}$=1.85 $\pm$ 0.10 at 100 K.  Therefore, the high temperature low energy fluctuations in Fe$_{1.057(7)}$Te are anisotropic in momentum, despite the tetragonal shape of the lattice and the equivalence of the $a$ and $b$ directions.  However, these fluctuations centered around the ($\pi$, 0) position do preserve the C$_{4}$ symmetry of the lattice and should be distinguished from the ``nematic" phase fluctuations identified in the ``122" pnictides at high temperatures.~\cite{Lu14:345,Fernandes14:10}  The anisotropy around the ($\pi$,0) position may reflect the underlying Fermi surface~\cite{Subedi08:78} as suggested to explain a similar anisotropy in the magnetic fluctuations in iron based pnictides.~\cite{Park10:82,Graser09:11,Lu14:345} 

\begin{figure}[t]
\includegraphics[width=9.3cm] {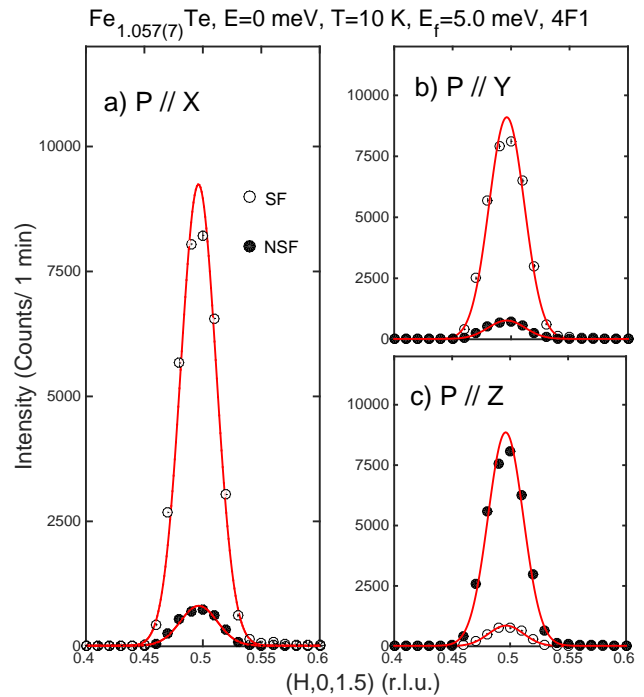}
\caption{\label{pol_elastic} Polarization analysis of the elastic (0.5, 0, 1.5) magnetic Bragg peak showing spin-flip (open circles) and non-spin-flip (filled circles) scattering with the incident beam of neutrons polarized along the X, Y, and Z directions as defined in the main text.  The peak in the non spin-flip channel in panel $(a)$ is the result of incomplete polarization of the neutron beam and is defined by the flipping ratio.} 
\end{figure}

We now investigate the polarization of the magnetic fluctuations as a function of temperature using polarized neutron scattering obtained at the 4F1 triple-axis spectrometer.  Figure \ref{pol_elastic} illustrates scans through the low temperature elastic magnetic Bragg peak at (0.5, 0, 1.5).  Spin-flip (open circles) and non spin-flip (filled circles) are illustrated for the neutron beam polarized along the X (defined as parallel $\vec{Q}$), Y (perpendicular to $\vec{Q}$, but within the horizontal (H0L) scattering plane), and Z (perpendicular to the $\vec{Q}$ and perpendicular to the horizontal scattering plane).  Panel $(a)$ shows that the dominant cross section is in the spin-flip channel, as expected for magnetic scattering, with the feed-through measured in the non-spin-flip channel the result of incomplete polarization characterized by the flipping ratio discussed above in the experimental section.   Scans with the polarization along Y indicate a strong spin-flip cross section indicating that the magnetic moment is oriented out of the scattering plane.  This is confirmed by scans with the neutron polarization oriented along Z which show a dominant cross section in the non-spin-flip channel.  Polarization analysis along the Y and Z directions confirm that the magnetic moments are aligned along the $b$ axis, perpendicular to the (H0L) scattering plane chosen for the 4F1 polarized experiments.   This result is consistent with previous powder diffraction and single crystal neutron diffraction reported for the iron deficient side of the Fe$_{1+x}$Te phase diagram.

\begin{figure}[t]
\includegraphics[width=9.0cm] {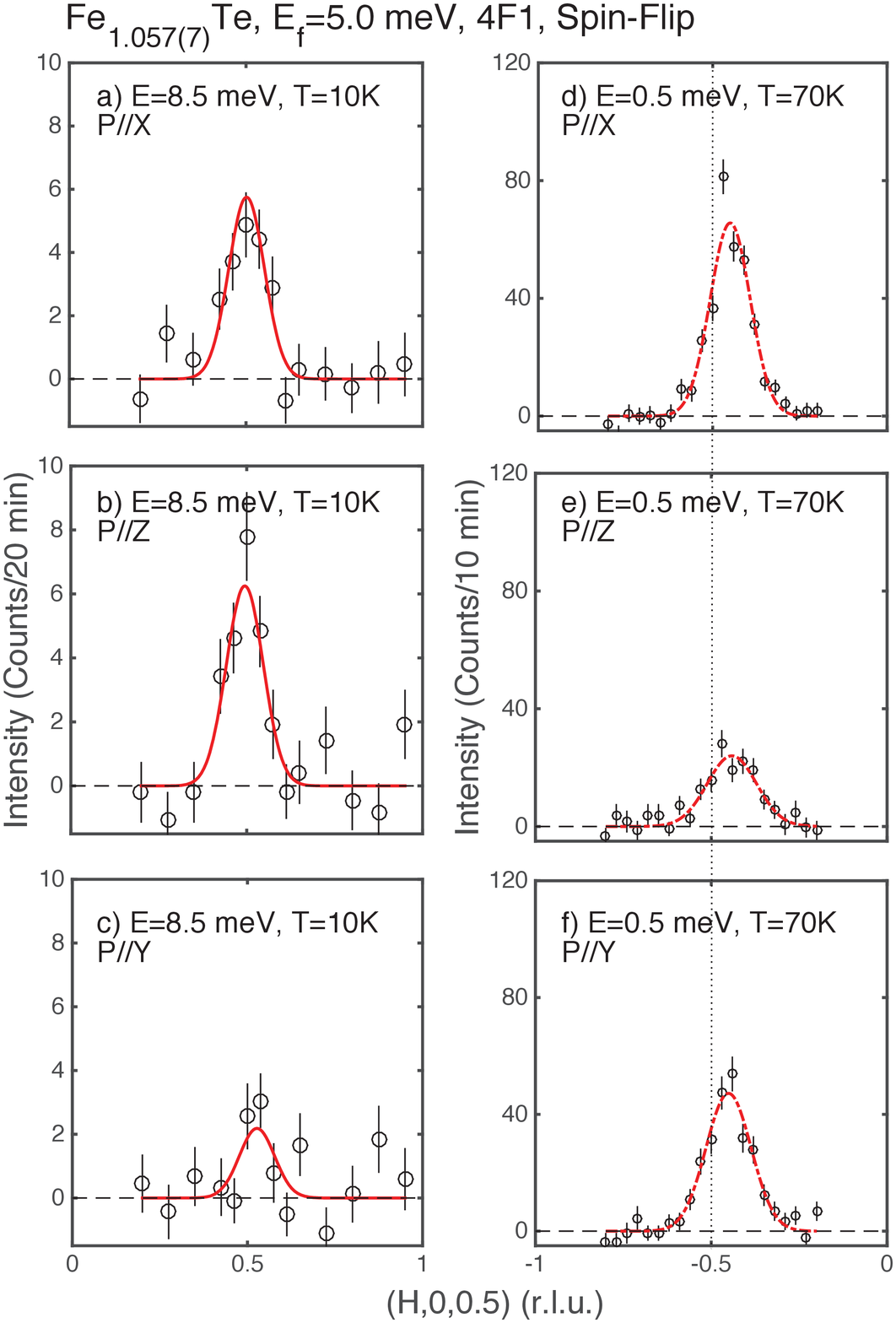}
\caption{\label{pol_cuts} Constant energy scans with polarization analysis at low temperatures at E=8.5 meV (panels $a-c$) and near the Neel transition at E=0.5 meV.  Given that the ordered magnetic moment is aligned along the $b$ axis, the polarized scans illustrated in $(a-c)$ show the dominant magnetic cross section is transverse to the magnetic moment direction.  This is contrasted with panels $(d-f)$ which illustrate a magnetic cross section predominately polarized along the $b$-axis.  The high temperature scattering also appears a $\vec{Q}_{0}$=($\sim$ 0.45,0,0.5) which is contrasted with the commensurate magnetic scattering at low temperatures (illustrated by the dotted line).}
\end{figure}

Having reviewed the magnetic structure at low temperatures with elastic neutron scattering with polarization analysis, we now discuss the polarization of the low temperature spin fluctuations.  The low temperature magnetic dynamics in Fe$_{1.057(7)}$Te are gapped for this particular iron concentration~\cite{Stock11:84,Stock14:90} as shown in Fig. \ref{constQ}.  In Fig. \ref{pol_cuts} panels $(a-c)$, we investigate the polarization of these fluctuations at an energy transfer of E=8.5 meV, above the energy gap.  Panel $(a)$ shows the total magnetic cross section as probed in the spin-flip channel with the neutron beam polarized along X.   Panel $(b)$ illustrates the same scan, but now with the neutron beam polarized along the Z direction (perpendicular to $\vec{Q}$ and the horizontal (H0L) scattering plane utilized on 4F1).  Given the geometry of the spectrometer and sample, this corresponds to the $b$ axis of the sample.  The intensity measured in this channel is, within error, equal to the total magnetic cross section measured in panel $(a)$ with the neutron beam polarized along X.  A small spin-flip cross section is measured with the beam polarized along Y.  This scan is sensitive to spin fluctuations along the $b$ axis of the material and parallel to the low temperature ordered magnetic moment direction.  Given the statistics, it is not clear if this is statistically significant given the flipping ratio.  The main result found in the polarization analysis in panels $(a-c)$ is that the dominant magnetic cross section at E=8.5 meV is transverse to the ordered magnetic moment direction at low temperatures.  We therefore conclude that the low energy spin fluctuations in Fe$_{1.057(7)}$Te are the result of localized spin fluctuations similar to spin-waves in an ordered antiferromagnet.  

Figure \ref{pol_cuts} $(d-f)$ show polarization analysis at E=0.5 meV of the low energy fluctuations at 70 K near the N\'eel temperature (T$_{N}$=75 K).     As noted previously~\cite{Parshall12:85}, these fluctuations are incommensurate at H$\sim$ 0.45 and this is highlighted by the vertical dashed line at the commensurate H=0.5 position in Fig. \ref{pol_cuts}.  Panel $(d)$ shows the total magnetic cross section with the neutron beam polarized along $\vec{Q}$, defined as the X direction.  Panel $(e)$ shows a weaker cross section corresponding to fluctuations perpendicular to the $b$ axis of the sample (the low temperature ordered magnetic moment direction), however, a larger cross section is found in panel $(f)$ with the Y-polarized neutrons.  This analysis suggests a dominant fraction of the neutron cross section at 70 K corresponding to fluctuations polarized along the $b$ axis which are longitudinal fluctuations parallel to the low temperature ordered magnetic moment.  

\begin{figure}[t]
\includegraphics[width=9.0cm] {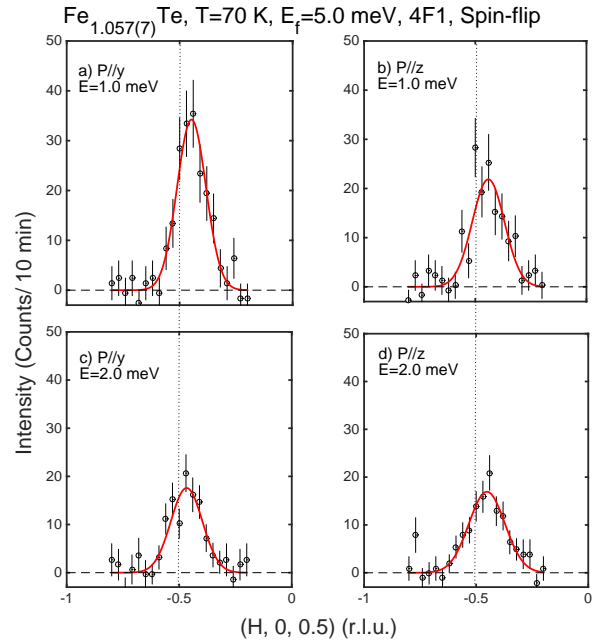}
\caption{\label{pol_energy} Polarization analysis of the incommensurate fluctuations near the Neel transition with neutrons polarized along the Y and Z directions as defined in the main text.  $(a)$ and $(b)$ illustrate an anisotropy in the fluctuations at 1.0 meV as evidenced by different intensities in the two channels.  At 2.0 meV (panels $c-d$) the fluctuations are isotropic with equal spectral weight in both polarization channels.  The vertical dashed line indicates the (0.5,0,0.5) position highlighting the fact that the high temperature spin fluctuations are incommensurate.}
\end{figure}

Figure \ref{pol_energy} illustrates the energy dependence of the incommensurate fluctuations critical to collinear Neel ordering.  Panels $(a,b)$ show polarization analysis at an energy transfer of 1.0 meV and panels $(c,d)$ at 2.0 meV.  The Y-polarized spin-flip channel is sensitive to fluctuations along the $b$ axis and the Z-polarized channel is sensitive to fluctuations transverse, or perpendicular, to $b$.  An anisotropy is observable at 1.0 meV, however at higher energy transfers of 2.0 meV (panels $c,d$), the excitations are isotropic within error with equal weight residing in the Y and Z polarized spin-flip channels.  This shows that the low energy incommensurate fluctuations are primarily longitudinal in nature, at higher energy transfers the fluctuations become more isotropic.

\begin{figure}[t]
\includegraphics[width=9.0cm] {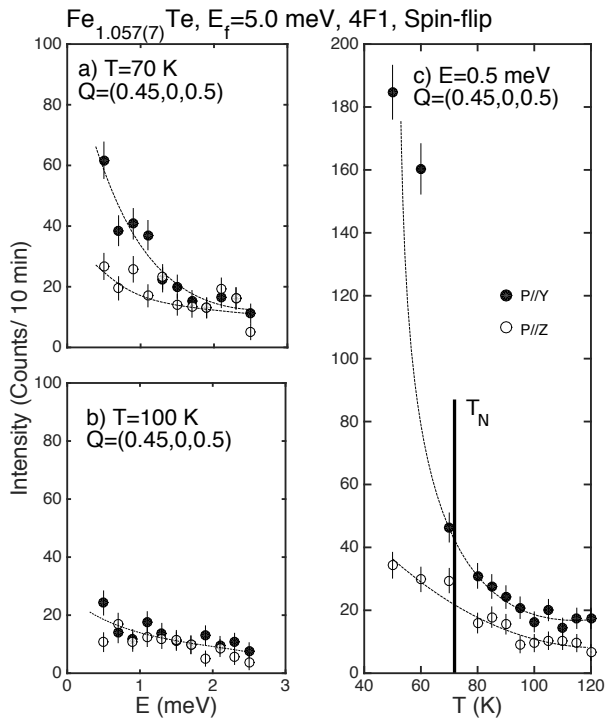}
\caption{\label{pol_temp}  The temperature, energy, and polarization dependence of the magnetic fluctuations at $\vec{Q}$=(0.45, 0 , 0.5).  The data was taken on the polarized cold triple-axis spectrometer 4F1.  Panel $(a)$ shows and energy scan with X and Y polarized neutrons at 70 K illustrating that the anisotropy develops between the two channels at low energy transfers.  $(b)$ shows the same constant-Q scan at 100 K illustrating that the fluctuations are isotropic, within error, at this temperature for all energy transfers investigated.  $(c)$ illustrates a temperature scan with E=0.5 meV and $\vec{Q}$=(0.45,0.5,0.5) for Y and Z polarized neutrons.  The anisotropy between the two channels develops near T$_{N}$.}
\end{figure}

Figure \ref{pol_temp} illustrates background corrected temperature and energy scans for the incommensurate magnetic fluctuations peaked at (0.45, 0, 0.5).  Panel $(a)$ and $(b)$ display constant momentum cuts.  At 70 K, near the N\'eel temperature, a significant difference develops between the Y and Z polarized channels at low energy transfers below $\sim$ 1 meV.    At higher temperatures of 100 K displayed in panel $(b)$, the two channels for neutrons polarized along Y and Z have equal intensities within error indicating isotropic fluctuations at all energy transfers studied.  This is expected for a paramagnet at temperatures well above the ordering temperature.   The temperature dependence of the magnetic fluctuations at E=0.5 meV and with $\vec{Q}$=(0.45, 0, 0.5) is displayed in panel $(c)$ where it is seen that a large difference between the spin-flip channels with Y and Z polarized neutrons is present near and below T$_{N}$.  At high temperatures the two channels are equal within error.

The polarized neutron scattering results demonstrate anisotropic spin fluctuations which develop near T$_{N}$ in Fe$_{1.057(7)}$Te.  This is evidenced in the difference seen between the Y and Z polarization channels in Figs. \ref{pol_cuts}, \ref{pol_energy}, and \ref{pol_temp} discussed above.  If the magnetic fluctuations were isotropic, the intensity in these two spin-flip channels would be equal and ${1\over 2}$ the intensity when the neutron beam is polarized along $\vec{Q}$ as observed in magnets in the paramagnetic region at high temperatures as shown in Refs. \onlinecite{Wicksted84:30,Ishikawa85:31}.   These anisotropic fluctuations are preferentially polarized along the $b$ axis which is parallel to the low temperature ordered magnetic moment.  However, these fluctuations are located at an incommensurate wave vector of $\vec{q}_{0}$=($\sim$ 0.45, 0, 0.5) and are distinct from the low temperature commensurate magnetic order and the fluctuations associated with this order which occurs at (0.5, 0, 0.5).  This indicates that these high temperature fluctuations are associated with a competing phases.  The polarization and also the wavevector are the same as the collinear spin-density wave reviewed above for Fe$_{0.124(5)}$Te.  We therefore conclude that this magnetic density wave phase competes with collinear and commensurate order in the Fe$_{1+x}$Te phase diagram.

\section{The magnetic structure in x=0.141(5) - Helical magnetism and ordered interstitial iron sites}

Having discussed the competition between and localized collinear magnetism and spin density wave phase in iron deficient Fe$_{1+x}$Te for $x$ less than $\sim$ 0.12, we now discuss single crystal neutron diffraction for large concentrations of interstitial iron where helical magnetic order has been previously observed.   

Owing to the presence of ferromagnetic iron oxide near the surface of the single crystal, polarized experiments on large interstitial iron concentrations were not successful.  Therefore we pursued single crystal unpolarized measurements on HB-3A (Oak Ridge).  

Large concentrations of interstitial iron have been found to result in semiconducting or poorly metallic behavior over a broad temperature range.~\cite{Rodriguez13:88}  Further transport studies on superconducting samples of Fe$_{1+x}$Te$_{1-y}$(Se,S)$_{y}$ found evidence for interstitial iron even causing charge localization.~\cite{Liu09:80}  Here we use single crystal neutron diffraction to investigate the magnetism on the interstitial iron site in Fe$_{1.141(5)}$Te.

\begin{figure}[t]
\includegraphics[width=8.5cm] {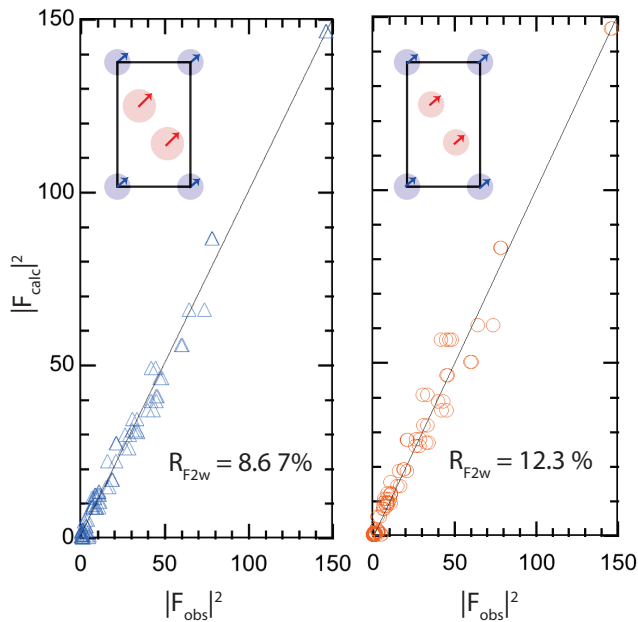}
\caption{\label{refinement} Results from single magnetic structure refinement on a single crystal of Fe$_{1.141(5)}$Te performed on the HB-3A diffractometer.  $(a)$ shows the results of a refinement where the interstitial iron moment sizes was allowed to vary while $(b)$ shows a refinement where they were constrained to be equal.  For model $(a)$ the refined size of the Fe moment in the FeTe layers is 2.01(2) $\mu_{B}$ and 3.5(5) $\mu_{B}$ on the interstitial site.  In model $(b)$, the iron moments in both the FeTe layers and interstitial sites were constrained to be equal giving a refined moment size of 2.1(1) $\mu_{B}$.}
\end{figure}

Results of a single crystal refinement for Fe$_{1.141(5)}$Te is illustrated in Fig. \ref{refinement} which plots $|F_{cal}|^{2}$ as a function of $|F_{obs}|^{2}$ with the $R$-factor listed for each fit.  This particular concentration of interstitial iron is placed beyond the Lifshitz point separating collinear and helical magnetism.  Two models are shown in Fig. \ref{refinement}, the first where the interstitial iron moment size was allowed to vary independently of the moment size in the FeTe layers and the second where both were constrained to be equal.   The first model (panel $a$) refines to 2.01(2) $\mu_{B}$ and 3.5(5) $\mu_{B}$ respectively for iron in the FeTe layers and interstitial sites respectively.   The constrained model (panel $b$) refines to 2.1(1) $\mu_{B}$.  

The refined helical magnetic structure in Fe$_{1.141(5)}$Te is different to the collinear phase found for smaller interstitial iron concentrations.  It is also different to the helical phase in FeAs~\cite{Rodriguez11:2,Segawa09:78,Selte69:23} which displays a noncollinear spin density wave with the spin amplitude along the $b$ axis direction larger than the $a$ direction.  The magnetic iron moments refine to a uniform helical magnetic structure.  The refinement also illustrates that the interstitial sites are fully ordered with a moment size that is comparable, and larger within error, to ordered magnetic moments within the FeTe layers. We note that while powder diffraction results indicated a substantial magnetic moment on the interstitial site, the single crystal results presented here confirm this result along with the fact that the interstitial site follows the same magnetic structure as the FeTe layers.  

\section{$x$=0.124(5)- Spin density wave and search for charge wave}

\begin{figure}[t]
\includegraphics[width=8.8cm] {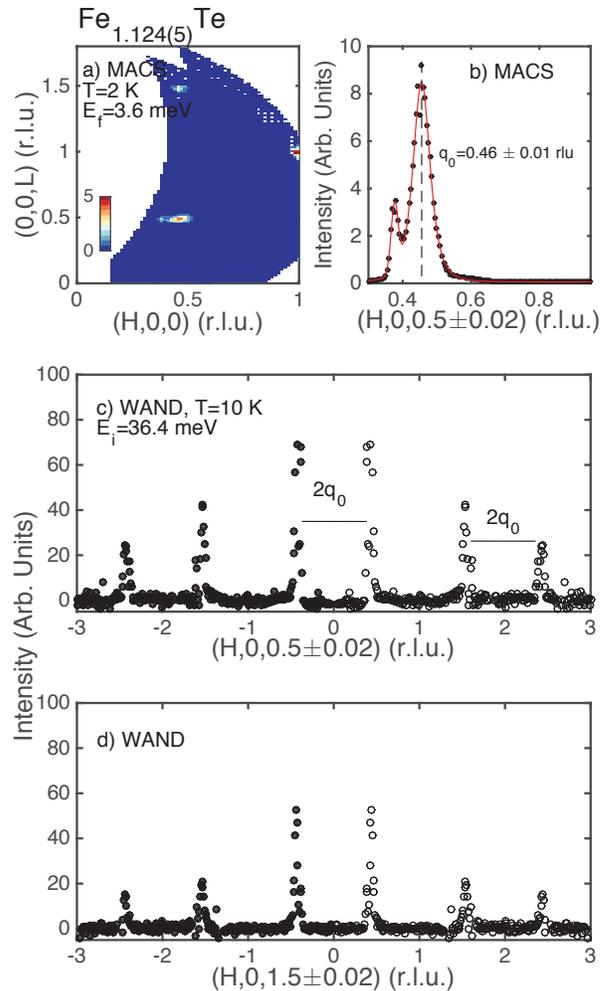}
\caption{\label{prop_vector} $(a)$  Elastic energy slice taken on the MACS spectrometer.  $(b)$ displays a cut with L=0.5 illustrating the incommensurate wavevector.  The smaller peak at lower H results from the inclusion of a magnetic helical phase with a larger interstitial iron concentration.~\cite{Rodriguez11:84}  $(c-d)$ show cuts taken on the WAND diffractometer illustrating the propagation vector.  The filled circles are symmetrized data and displayed to illustrate the relative positions of the magnetic scattering.}
\end{figure}

We now discuss the magnetic properties at the border between collinear and helical order in the Fe$_{1+x}$Te phase diagram by presenting neutron diffraction data on a single crystal of Fe$_{1.124(5)}$Te.

Measurements of the elastic neutron cross section at the border between collinear antiferromagnetism and the helical phase were done using a single crystal of Fe$_{1.124(5)}$Te.   Figure \ref{prop_vector} $(a)$ illustrates a constant energy slice at the elastic position taken on the MACS cold triple axis spectrometer at 2 K.  Panel $(b)$ displays a cut along the H direction illustrating the incommensurate wavevector at $q_{0}$=0.46 $\pm$ 0.01 along the H direction based on fits to a Lorentzian squared lineshape.  Within experimental error, the peak is commensurate along L being positioned at L=0.5 and no observable evidence of second harmonics at 2$q_{0}$ are observable in the data (within 2\% of the peak height at $q_{0}$=0.46 $\pm$ 0.01).  As discussed previously in Ref. \onlinecite{Rodriguez11:84}, the static magnetism corresponding to this peak is short-range along the $a$ axis evidenced by a broader than resolution lineshape along the H direction.  The lineshape is resolution limited along $c$ corresponding to long range order along L.  An analysis based on polarized neutrons found that the magnetic structure is polarized along the $b$ axis in contrast to the helical order for Fe$_{1+x}$Te samples on the iron rich side of the phase diagram.  The magnetic structure at $x$=0.124(5) therefore corresponds to a collinear spin-density wave phase.  A second peak is observed at a lower $q$ position and as discussed in Refs. \onlinecite{Rodriguez11:84,Rodriguez13:88} based on a polarized neutron analysis, this corresponds to a small inclusion of a helical phase with a larger interstitial iron concentration.  

As discussed above in the introduction, there have been several reports of magnetism at this incommensurate wave vector, even in superconducting samples doped with Se.  However, it is not clear from the limited momentum range if the peak is incommensurate with respect to the nuclear positions or to the antiferromagnetic H=0.5 point as might be expected based on analogies with cuprates.  Figure \ref{prop_vector}, panels $(c)$ and $(d)$, show cuts at L=0.5 and L=1.5 taken on WAND where the combined thermal neutron wavelengths and broad detector coverage allow us to study the magnetism over a broad range of momentum transfer.  The cuts prove the result reported previously that the propagation vector is $q_{0}$=0.46 $\pm$ 0.01 and is incommensurate with respect to the nuclear positions.  This contrasts with some studies that have stated that the propagation vector is taken as (0.5-$\delta$,0,0.5)~\cite{Wen12:86} and is only clear in the current data set given the broad momentum coverage afforded by the WAND diffractometer.   

\begin{figure}[t]
\includegraphics[width=9.5cm] {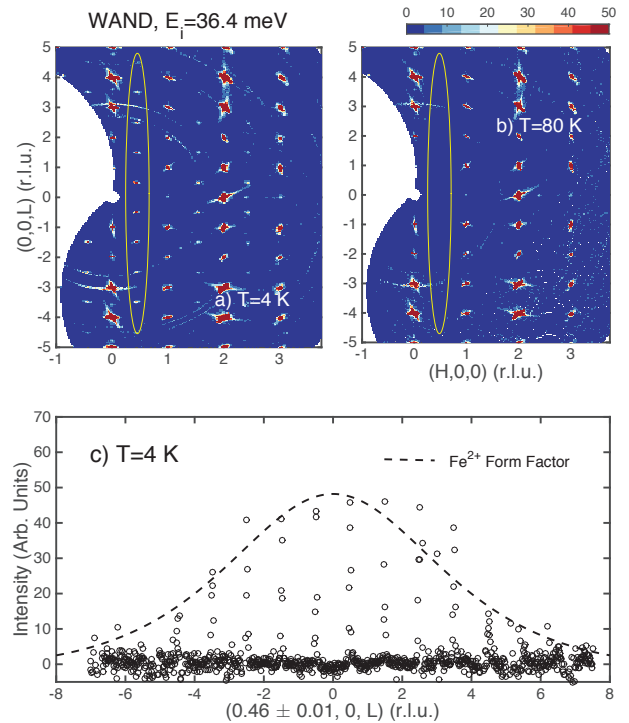}
\caption{\label{wand_mag} Elastic momentum slices obtained from the WAND diffractometer.  $(a)$ and $(b)$ show scans obtained at 4 K and 80 K in the spin density wave and paramagnetic phase respectively. $(c)$ shows a cut along the L direction with H=0.46 $\pm$ 0.01 r.l.u. showing the half integer commensurate nature of this scattering and also that it follows the expected decay of intensity based on the Fe$^{2+}$ form factor.  Backgrounds with an empty can have been obtained at both temperatures and subtracted.}
\end{figure}

Figure \ref{wand_mag} illustrates an extensive reciprocal space map at 4 K (in the magnetically ordered state) and also at high temperatures of 80 K where Fe$_{1.124(5)}$Te is paramagnetic.   A series of magnetic superlattice peaks are clearly observed at H$\sim$ 0.46, 1.54 and 2.45 r.l.u. at T=4 K, but absent at high temperatures of 80 K, confirming the magnetic origin.  This is highlighted by the yellow ellipse at H $\sim$ 0.46 at both temperatures.  Panel $(c)$ plots an L scan at H=0.46 $\pm$ 0.01 r.l.u. showing that the magnetic peaks appear at the commensurate half integer positions along L and also that the intensity decays with the expected Fe$^{2+}$ form factor.  This is consistent with dipolar selection rules for the intensity based on localized magnetic moments pointing along the $b$ axis.

Our WAND results are not consistent with suggestions of antiphase boundaries separating locally ordered collinear states.~\cite{Mazin09:5}  While such a structure can produce scattering at incommensurate positions, as observed in stripe phases of nickelates~\cite{Tranquada96:54} and also cuprates~\cite{Waki00:61,Waki99:60,Stock04:69,Stock10:82}, it fails to model both the incommensurate wavevector and the lack of higher harmonics that would be associated with a sharp uniaxial boundary.  It has recently been proposed that the structure maybe understood in terms of solitons~\cite{Materne15:115}, however the magnetic structure proposed would produce a $c$ and $a$ axis component to the scattering in our previous polarized neutron diffraction studies of this compound.  This contradicts the data which is consistent with a component only along the $c$-axis.  We therefore conclude that the short-range static antiferromagnetism observed near interstitial iron concentrations of $x\sim$ 0.12 is more consistent with a spin density wave where the magnitude of the spin varies along the direction of propagation.  

\begin{figure}[t]
\includegraphics[width=9.5cm] {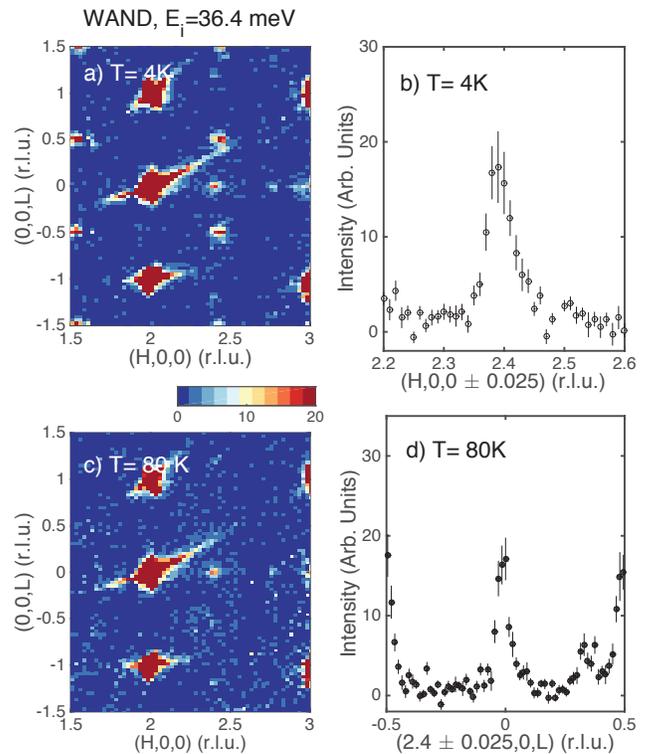}
\caption{\label{wand_nuc} Momentum slices from WAND of the ``superlattice" nuclear peak observed at (2.39 $\pm$ 0.05, 0, 0$\pm$ 0.05).  $(a)$ and $(c)$ show the peak to be present at 4 and 80 K, in the spin density and paramagnetic phase. $(b)$ and $(d)$ show cuts through the peak establishing its position in momentum. The lack of temperature dependence and inconsistency in the position near different nuclear Bragg peaks leads to believe this peak is spurious.}
\end{figure}

Using the wide momentum coverage on WAND, we have also searched for any charge density wave that may accompany this spin density wave.  The single crystal momentum maps in Fig. \ref{wand_mag} $(a)$ and $(b)$ display no observable superlattice peaks that may be associated with a charge density wave.  A small peak is observed near (2.39 $\pm$ 0.05, 0, 0$\pm$ 0.05), however this peak is present at both 4 K and 80 K and is not observable near any other primary nuclear Bragg peak measured in our reciprocal space mapping.  The peak is illustrated in Fig. \ref{wand_nuc}.  While a coupling between strain or charge and magnetism is expected,~\cite{Turner09:80} as displayed in Cr metal~\cite{Fawcett88:60}, the charge density wave peak intensity is in proportion to the spin density wave and should be at harmonic of the primary wavevector which is not the case here.~\cite{Pynn76:13,Kotani76:41,Kotani78:44}   

The lack of consistency with what has been discussed in relation to coupled spin and charge density waves and the wave vector imply that this peak is not associated with a charge density wave.  From an experimental viewpoint, the peak is also suspicious given the large tails from the (200) peak likely originating from strong bragg scattering feeding through the collimators.~\cite{Shirane:book}  A similar structure can be seen near (004) which is also a strong nuclear Bragg peak.  The inconsistency between different Bragg positions and also the correlation with strong nuclear Bragg lead us to conclude that the peak is likely spurious.  These small weak peaks are likely due to secondary scattering from the aluminium window located on the multi wire detector system.

\section{Conclusions and Discussion}

We have presented a study of the magnetic structures and fluctuations near the Lifshitz point in the Fe$_{1+x}$Te phase diagram using neutron diffraction, inelastic scattering, and neutron polarization analysis.  We have identified an incommensurate spin density wave which competes with collinear ``double-stripe" magnetism and is stabilized over a narrow range of interstitial iron concentrations.  These incommensurate fluctuations are highly anisotropic being longitudinally polarized along $b$ and also anisotropic in the dynamical correlation lengths as evidenced by differing widths along H and K.  The anisotropy reflects stronger correlations along the $a$ direction and weaker correlations along $b$.  Such anisotropy was also observed in superconducting FeTe$_{1-x}$Se$_{x}$~\cite{Xu16:93} and modelled with correlations reflecting a such anisotropy, but tilted to reflect that the correlations were centered around the ($\pi$, $\pi$) position instead of the  ($\pi$, 0) position found here.  These incommensurate correlations are observed to compete with collinear magnetic order for interstitial iron concentrations less than $x\sim$0.12. For larger interstitial iron concentrations this competition between incommensurate spin density fluctuations and collinear order is replaced by robust helical magnetic order.  Our studies of helically ordered samples have not observed any evidence of incommensurate order at H$\sim$ 0.45 even above T$_{N}$.~\cite{Rodriguez13:88}  Based on this, we speculate that this incommensurate order competes with collinear magnetic order while helical magnetism is a robust feature of the Fe$_{1+x}$Te phase diagram.

In previous studies, we have related the temperature dependence of these fluctuations to the resistivity with the suggestion that they are the origin of the semiconducting properties above T$_{N}$ in the collinear magnetically ordered part of the Fe$_{1+x}$Te phase diagram.~\cite{Rodriguez13:88}  The results are also confirmed by measurements under pressure where the structure can be tuned in a similar manner to doping interstitial iron.~\cite{Okada09:78}  The temperature range also coincides with where optical conductivity observes strong fluctuations in the terahertz regime~\cite{Homes10:81} which is the same energy and temperature range where we observe highly anisotropic spin density wave fluctuations. While these correlations are dynamic for the iron deficient region of the Fe$_{1+x}$Te phase diagram, static spin density magnetism is stabilized for a narrow region of $x$ $\sim$ 0.12.  

Localized models of the magnetism in Fe$_{1+x}$Te have been very successful in predicting the two dominant magnetic and structure phases - the collinear magnetic phase at small interstitial iron $x$ and the helical magnetic phase at large interstitial iron concentrations.~\cite{Chen13:88}  However, to our knowledge, these models have not predicted the density wave phase we observe that competes with collinear magnetism and is stabilized for a narrow range of interstitial iron concentrations near $x\sim0.12$.  These models also do not account for the anisotropic magnetic fluctuations at temperatures above T$_{N}$ which are incommensurate, polarized along $b$, and correlated anisotropically in momentum.  The existence of an instability to orbital order could help explain the origin of a large anisotropy in the paramagnetic fluctuations illustrated above in Fe$_{1.05(7)7)}$Te.  This extra order parameter present at high temperatures, and the concomitant structural and magnetic transition at T$_{N}$, has been used to explain the resistivity anomaly observed on the interstitial iron poor side of the Fe$_{1+x}$Te phase diagram.~\cite{Bishop16:117}

The presence of an orbital degree of freedom has also been implicated in understanding the magnetic correlations and phase transitions in iron based systems.~\cite{Khal14:90,Chen09:80}  The Hunds rule coupling has been cited as the origin of the strong electronic correlations~\cite{Yin10:105,Liang12:109,Lanata13:87,Bascones12:86} along with multi orbital models.~\cite{Kubo07:75,Zhang09:79,Singh15:1}  where localized magnetism exists on some orbitals while others are more itinerant.~\cite{Ducatman14:90}  The presence of an delocalized orbital degree of freedom may also support recent high energy neutron inelastic scattering measurements which find a deficit of spectral weight and also a considerable energy dampening of the excitations at high energy transfer.~\cite{Stock14:90}  

In summary, we have reported the competition between a spin density wave phase with localized collinear and helical magnetism for interstitial iron concentrations near the Lifshitz point at $x\sim$ 0.12.   Using neutron diffraction, we report on the magnetic structure near this point.  Polarized neutron inelastic scattering for concentrations less than $x\sim$ 0.12 observe a competition between localized and longitudinal polarized spin fluctuations.  For larger interstitial iron concentrations, this is replaced by robust helical magnetic order. Based on these results, we suggest the presence of a spin density wave which competes with ordered antiferromagnetism in the Fe$_{1+x}$Te phase diagram.

This work was funded by the Carnegie Trust for the Universities of Scotland, the Royal Society of Edinburgh, the EPSRC, and through the National Science Foundation (Grant No. DMR-09447720). Research at Oak Ridge National Laboratory's HFIR was sponsored by the Scientific User Facilities Division, Office of Basic Energy Sciences, U. S. Department of Energy.


%

\end{document}